\documentclass[preprint,12pt]{elsarticle}
\usepackage[T1]{fontenc}
\usepackage[latin1]{inputenc}
\usepackage{amsmath,amssymb,amsthm}
\usepackage{graphicx}
\usepackage{graphics}
\usepackage{multirow}
\usepackage{multicol}
\usepackage{amsmath}
\usepackage{amsfonts}
\usepackage{amssymb}
\usepackage{epstopdf}
\usepackage{epsfig}
\usepackage{subfigure}
\usepackage{doi}
\usepackage{hyperref}

\def\be{\begin{equation}}
\def\ee{\end{equation}}
\def\bea{\begin{eqnarray}}
\def\eea{\end{eqnarray}}

\begin{document}

\begin{center}
{\Large \bf Area distribution of two-dimensional random walks \\  and non Hermitian Hofstadter quantum mechanics}\\[0.5cm]

{\large \bf Sergey Matveenko}\footnote{matveen@landau.ac.ru}\\[0.1cm]
Landau Institute for Theoretical Physics,  Kosygina Str. 2,
119334, Moscow, Russia \\ and \\Laboratoire de Physique Th\'eorique et Mod\`eles
Statistiques, 91405 Orsay, France
\\[0.4cm]
{\large \bf St\'ephane Ouvry}\footnote{stephane.ouvry@u-psud.fr}\\[0.1cm]
 Laboratoire de Physique Th\'eorique et Mod\`eles
Statistiques\footnote{Unit\'e Mixte de Recherche CNRS-Paris Sud, UMR 8626}\\
91405 Orsay, France
\\[0.2cm]

\today

\end{center}

\vskip 0.5cm
\centerline{\large \bf Abstract}
\vskip 0.2cm
When   random walks on a square lattice
   are biased horizontally to move  solely to the right, the probability distribution of their algebraic area  can be exactly  obtained \cite{nous}. We explicitly map this biased classical random system on a non hermitian  Hofstadter-like quantum model where a charged particle on a square lattice coupled to a perpendicular magnetic field hopps   only to the right. In the commensurate case when the magnetic flux per unit cell is rational, an exact solution of the quantum model is obtained. 
Periodicity  on the lattice allows to relate   traces of the $N^{\rm th}$ power of the  Hamiltonian to   probability distribution generating functions of  biased walks of length $N$. 


\section{Introduction}

It is well-known that the probability distribution for the algebraic area 
enclosed by closed random walks  on a two-dimensional square lattice is related to the Hofstadter model \cite{Hofstadter}
of an electron hopping on a  square lattice and coupled to a  perpendicular magnetic field. The generating function of the algebraic area probability distribution of  walks of length $N$ is  formally identified with the
trace of the $N^{\rm th}$ power of the Hofstadter  Hamiltonian.   
This mapping has been used  in \cite{Bell} to recover   asymptotically Levy's law \cite{Levy} and its first $1/N^2$ correction.
Here we consider random walks biased  horizontally to move only to the right, a geometry which is intermediate between 2 and 1 dimensions.  For such biased walks  the probability distribution of their algebraic area  has been exactly  obtained in \cite{nous}. We are going to relate the generating function of the algebraic area probability distribution  of biased walks of length $N$  to the trace of the $N^{\rm th}$ power of the Hamiltonian of a non hermitian Hofstadter-like quantum model. This situation is reminiscent of other  biased classical systems mapped on  non hermitian quantum models, as  for example the TASEP models \cite{Tasep}  and their corresponding  non hermitian quantum spin chains. It would  certainly be rewarding to look at possible physical interpretations, if any,  of the non hermitian Hofstadter quantum mechanics discussed here, in particular in relation to the quantum Hall effect. Note that in the  asymptotic  limit $N\to\infty$  one expects  to recover the probability distribution of the algebraic area under a 1d random curve, a distribution which can be easily obtained by more direct means \cite{inpreparation}.


\section{Algebraic area probability distribution generating function  for biased random walks on a square lattice}

The generating function for the algebraic area probability distribution  of closed walks of length $N$ -in the case of closed walks $N$ is necesseraly even- is defined as
\be\label{fine}
Z_N({\bf q}) \equiv  \sum_{A=-\infty}^\infty C_N(A)\, {\bf q}^A 
\ee
where $C_N(A)$ is the number, among the ${N\choose {N/2}}^2$, of closed walks    whose algebraic area is $A$. The mapping to the Hofstadter model is obtained by setting ${\bf q}=e^{i\gamma}$ where $\gamma = 2 \pi {\boldsymbol{\Phi}}/{\boldsymbol{\Phi}}_0$ is the flux $ {\boldsymbol{\Phi}}$ through the unit cell   in unit of the flux quantum ${\boldsymbol{\Phi}}_0 = hc/e$.

More generally, the algebraic area of an opened random walk with $M_1$ steps  right, $M_2$ steps  left, $L_1$ steps up and $L_2$ steps down
 can be defined as   those of the closed walk obtained by adding to the end point of the opened walk   a vertical path linking it   to the horizontal axis where it started from  and then adding a horizontal path back to its starting point. For such walks  the generating function $Z_{M_1,M_2,L_1,L_2}({\bf q})$ for their algebraic area probability distribution obeys  the 
recurrence relation \cite{nous,Beguin}
 \bea
Z_{M_1,M_2,L_1,L_2}({\bf q}) & = &  Z_{M_1,M_2,L_1-1,L_2}({\bf q})+ Z_{M_1,M_2,L_1,L_2-1}({\bf q}) \nonumber \\
&& {} + {\bf q}^{L_2-L_1}Z_{M_1-1,M_2,L_1,L_2}({\bf q}) + {\bf q}^{L_1-L_2}Z_{M_1,M_2-1,L_1,L_2}({\bf q}) \;,
\label{recbis}
\eea
with the initial condition $Z_{0,0,0,0}({\bf q}) = 1$.

\noindent Let us recall that if $x$ and $y$ are two operators satisfying $xy={\bf q}yx$   
 the  ${\bf q}$-binomial theorem   \cite{Andrews}
\be
(x + y)^{N} = \sum_{M=0}^{N} {N \choose M}_{\bf q} y^{N-M} x^M \;,
\label{qbin}
\ee
involves the ${\bf q}$-binomial coefficient
\be  
{N\choose M}_{\bf q}  \equiv \frac{[N]_{\bf q}!}{[M]_{\bf q}! [N-M]_{\bf q}!} \;,
\ee
and the {\bf q}-factorial
\be
[L]_{\bf q}!  =  \prod_{i=1}^{L}{1-{\bf q}^{i}\over 1-{\bf q}}
 =  1 (1+{\bf q}) (1+{\bf q}+{\bf q}^2) \cdots (1 + {\bf q} + \ldots + {\bf q}^{L-1}) \;.
\ee
Here formally one has  a generalized ${\bf q}$-binomial theorem with four addends
\be
(x + y + x^{-1} + y^{-1})^N =
\sum_{\stackrel{\scriptstyle M_1,M_2,L_1,L_2}{M_1+M_2+L_1+L_2 = N}}
Z_{M_1,M_2,L_1,L_2}({\bf q}) y^{-L_1}y^{L_2} x^{M_1} x^{-M_2} 
\label{qbingen}
\ee 
where the unknown $Z_{M_1,M_2,L_1,L_2}({\bf q})$'s obey  (\ref{recbis}).

An exact solution of (\ref{recbis}) for  random walks
 biased on the horizontal axis  to move solely to the right, i.e. $M_2=0$, has been obtained in \cite{nous}
\be\label{solut}
Z_{M_1,0,L_1,L_2}({\bf q}) = \sum_{k=0}^{\min(L_1,L_2)} \left[ {M_1 + L_1 + L_2 \choose k} - {M_1 + L_1 + L_2 \choose k-1} \right]
{M_1 + L_1-k \choose M_1}_{\frac{1}{{\bf q}}} {M_1 + L_2-k \choose M_1}_{{{\bf q}}}
\ee
such that a {\bf q}-binomial theorem for 3 addends  holds
\be
(x + y + y^{-1})^N =
\sum_{\stackrel{\scriptstyle M_1,L_1,L_2}{M_1+L_1+L_2 = N}}
 Z_{M_1,0,L_1,L_2}({\bf q})y^{-L_1}  y^{L_2} x^{M_1} \;.
\label{qbinter}
\ee
It is indeed not difficult to prove that (\ref{solut}) is the solution of (\ref{recbis}) when $M_2=0$.

\noindent One can go a step further and  "close" such biased walks   by enforcing them to return after $N$ steps to the  horizontal axis  they started from, i.e. $L_1=L_2$. It means that 
if one sets $M_1=M$, and so $L_1=L_2={N-M\over 2}$, the generating function for their algebraic area -in the sense  defined above-  is  then
\be\label{solutbis}
 Z_{M,0,{N-M\over 2},{N-M\over 2}}({\bf q})= \sum_{k=0}^{{N-M\over 2}} \left[ {N \choose k} - {N \choose k-1} \right]
{{N+M\over 2}-k \choose M}_{\frac{1}{{\bf q}}} {{N+M\over 2}-k \choose M}_{\bf q}\equiv Z_{N,M}({\bf q})\;. 
\ee
$ Z_{N,M}({\bf q})$ is by construction real.

\section{Random walks counting and non hermitian Hofstadter-like quantum mechanics}\label{mash}
\subsection{Random walks counting}\label{just}
The question we would like to address is : what the exact expression of $ Z_{N,M}({\bf q})$ in (\ref{solutbis}) can tell us on a corresponding Hofstadter-like model in quantum mechanics ? Recall  that in the Hofstadter case the counting of closed random walks of length $N$ directly follows from (\ref{fine}) by setting ${\bf q}=1$
\be Z_N(1)=\sum_{M=0}^{N/2}{N!\over M!^2({N-2M\over 2})!^2} ={N\choose N/2}^2=({1\over 2 \pi})^2\int_{-\pi}^{\pi}\int_{-\pi}^{\pi}(2\cos k_x+2\cos k_y)^Ndk_xdk_y\label{hof}\ee
The quantum spectrum in the RHS of (\ref{hof})  corresponds to the
  tight-binding  Hamiltonian 
\be H=T_x+T_x^{-1}+T_y+T_y^{-1}\ee
where $T_x=e^{ip_x/\hbar}$ and $T_y=e^{ip_y/\hbar}$ for a unit lattice step with Bloch
 eigenstates 
\be \label{bloch} e^{ik_xx}e^{ik_yy}\ee
where both $k_x$ and $k_y$  are in the interval $[-\pi,\pi]$. 
 The eigenenergies  are indeed
\be e^{ik_x}+e^{-ik_x}+e^{ik_y}+e^{-ik_y}=2\cos k_x+2\cos k_y\ee
as in the RHS of (\ref{hof}).

\noindent Similarly, in (\ref{solutbis}),  $Z_{N,M}(1)$  counts   the number of biased random walks  of length $N$ with $M$ steps to the right and ${N-M\over 2}$ steps up and  down

\be\label{solutter}
  Z_{N,M}(1)= {N!\over M!({N-M\over 2})!^2}\ee 
 For a given $N$, the number of all possible such walks is, 
 if $N$ is even,  
\be\label{solutquar}
\sum_{M=0, M {\;\rm even}}^N Z_{N,M}(1) =\sum_{M=0, M {\;\rm even}}^N{N!\over M!({N-M\over 2})!^2}\ee
and if $N$ is odd,  
\be\label{solutquint}
\sum_{M=1, M {\;\rm odd}}^N Z_{N,M}(1)=\sum_{M=1, M {\;\rm odd}}^N{N!\over M!({N-M\over 2})!^2}\ee
Both countings are equal\footnote{and to $\, _2F_1\left(\frac{1-N}{2},-\frac{N}{2};1;4\right)$. Note 
the  $N\to\infty$ asymptotic  scaling 
\be \sum_{ M=0, M {\;\rm even\; or \; odd}  }^N  Z_{N,M}(1)\simeq\frac{3^{N+\frac{1}{2}}}{2 \sqrt{\pi N}}\ee
 indicating a situation intermediate between $2$ and $1$ dimensions.
}  to ${1\over 2 \pi}\int_{-\pi}^{\pi}(\pm 1+2\cos k_y)^Ndk_y$: more precisely
 when $N$ is even
\be \sum_{ M=0, M {\;\rm even}  }^N  Z_{N,M}(1)={1\over 2 \pi}\int_{-\pi}^{\pi}(\pm 1+2\cos k_y)^Ndk_y\;,\label{even}\ee
and  when $N$ is odd
\be \sum_{ M=1, M {\;\rm odd}  }^N  Z_{N,M}(1)=(\pm 1){1\over 2 \pi}\int_{-\pi}^{\pi}(\pm 1+2\cos k_y)^Ndk_y\;.\label{odd}\ee
 In the RHS of (\ref{even},\ref{odd})  the spectrum
\be \label{plusmoin}\pm 1+2\cos k_y\ee
 corresponds  again to a tight-binding-like Hamiltonian but with only right hoppings on the horizontal axis
\be H=T_x+T_y+T_y^{-1}\ee
Indeed the   eigenstates (\ref{bloch})
has now for eigenenergies

\be e^{ik_x}+e^{ik_y}+e^{-ik_y}\ee
 If one insists on  restricting the Hilbert space to a real spectrum then  either $k_x=0$ or $k_x=\pm\pi$ so that one ends  up with

\be 1+2\cos k_y \quad {\rm or}\quad  -1+2\cos k_y\ee
i.e. (\ref{plusmoin}).
\subsection{Non hermitian Hofstadter-like quantum mechanics}
\noindent It follows that if one now introduces a magnetic field perpendicular to the lattice the mapping at hand  should be between   the algebraic area probability distribution   of random walks biased  horizontally to move  only to the right  and a non hermitian "Hofstadter-like" quantum mechanics  with  only right hoppings.
In
 the Landau gauge the  quantum Hamiltonian is
\be H_{\gamma}=T_x+T_y+T_y^{-1}\label{h2}\ee
where $T_x$ and $T_y$
act on  a state   $\Psi_{m,n}$  at the lattice site $\{m,n\}$ as \cite{magnoper}
\be
T_x \Psi_{m, n} = \Psi _{m + 1,  n}, \quad T_y \Psi_{m, n}  = e^{i \gamma m} \Psi _{m , n + 1}
\ee
and obey  the  commutation relation
\be
T_x T_y = e^{-i\gamma} T_y T_x,
\ee
 where, as said above, $\gamma = 2 \pi {\boldsymbol{\Phi}}/{\boldsymbol{\Phi}}_0$ is the flux $ {\boldsymbol{\Phi}}$ through the unit cell   in unit of the flux quantum ${\boldsymbol{\Phi}}_0 = hc/e$. The Hamiltonian (\ref{h2}) is non hermitian, therefore it has complex  non-physical eigenvalues.

 Using
translational invariance in the $y$ direction   one sets  $\Psi_{m,n}=e^{in k_y }\Phi_m$ to get   the eigenenergy  equation
\be \Phi_{m+1}+2\cos(k_y+\gamma m)\Phi_{m}=E\Phi_{m}\label{eq}\ee 
which can be iterated, for any  $k\ge 0$,  to
 \be
 \Phi_{m+k} = \prod_{r= m}^{m+k-1 } (E -2\cos (k_y+\gamma r ))\Phi_{m}.
  \label{wf}
 \ee

\section{Commensurate case: exact solution }
In the commensurate case  where the magnetic flux per  plaquette  is  rational
\be
\gamma  = 2 \pi \frac{p}{q}
\ee
with relatively prime integers $p$ and $q$,
 the model can be   solved exactly.
The flux  being rational, one has a  Harper-like model \cite{Harper}  of   period $q$ on the lattice. Using this periodicity, 
the eigenfunctions  in the  periodic potential are such that   
$\Phi_{m + q} = e^{i q k_x } \Phi_m$. It follows from (\ref{wf}) that

 \be\label{bibi}\prod_{r=1}^q (E-2\cos(k_y+2\pi {p\over q}r ))=e^{iqk_x}\ee
 has to be satisfied.
The product in the LHS of (\ref{bibi}) is independent of the integer $p$ that from now on will be fixed to $1$.
(\ref{bibi}) can be easily solved with respect to $E$ thanks to the identity \cite{prudnikov}
\be
\prod_{r=1}^{q} (a^2 - 2 a b \cos\left(k_y + \frac{2\pi }{q}r\right) + b^2) = a^{2 q} - 2 (a b)^q \cos (q k_y) + b^{2q},
\label{pr}
\ee
valid for any $a, b, q$ and $k_y$.
  Using (\ref{pr}) with $a b = 1$  and $a^2+b^2\equiv E$, one  rewrites the product  in the LHS of (\ref{bibi})   as
  \be\label{zero}\prod_{r=1}^q (E-2\cos(k_y+2\pi {p\over q}r ))=P_{q}(E)- 2\cos(q k_y)\ee
 where  $P_{q}$ is a Chebyshev polynomials of the first kind \cite{abr}   of degree $q$. 
Some examples of Chebyshev polynomials   are shown in Fig.~1 and Fig.~2.
 \begin{figure}[tbph]
  \begin{center}
\includegraphics[width=3.0in]{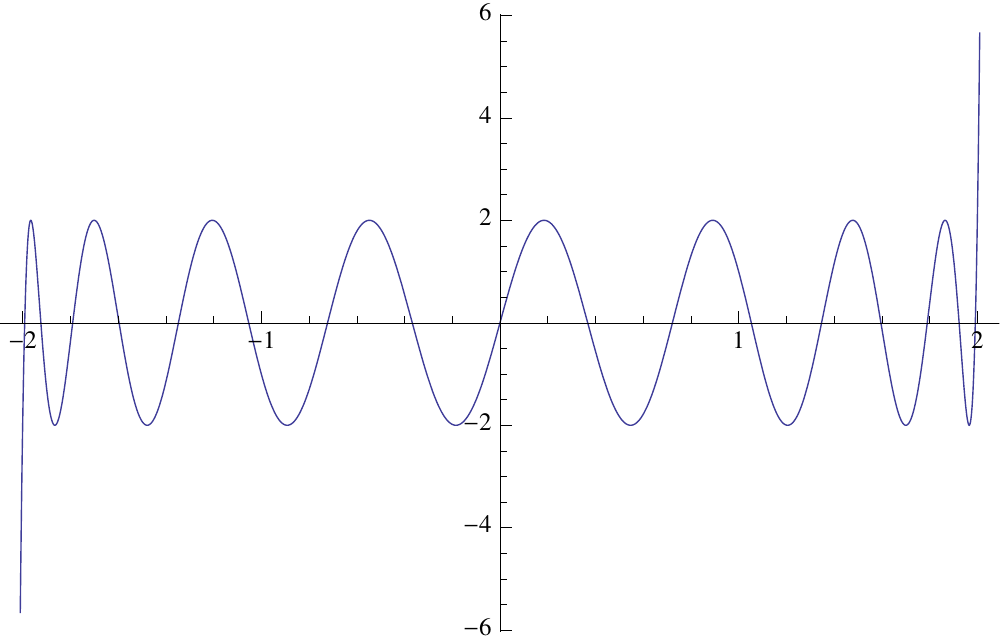}
\caption{  $P_{q=17}(E)$ }\label{fig01}
 \end{center}
\end{figure}

 \begin{figure}[tbph]
  \begin{center}
\includegraphics[width=3.0in]{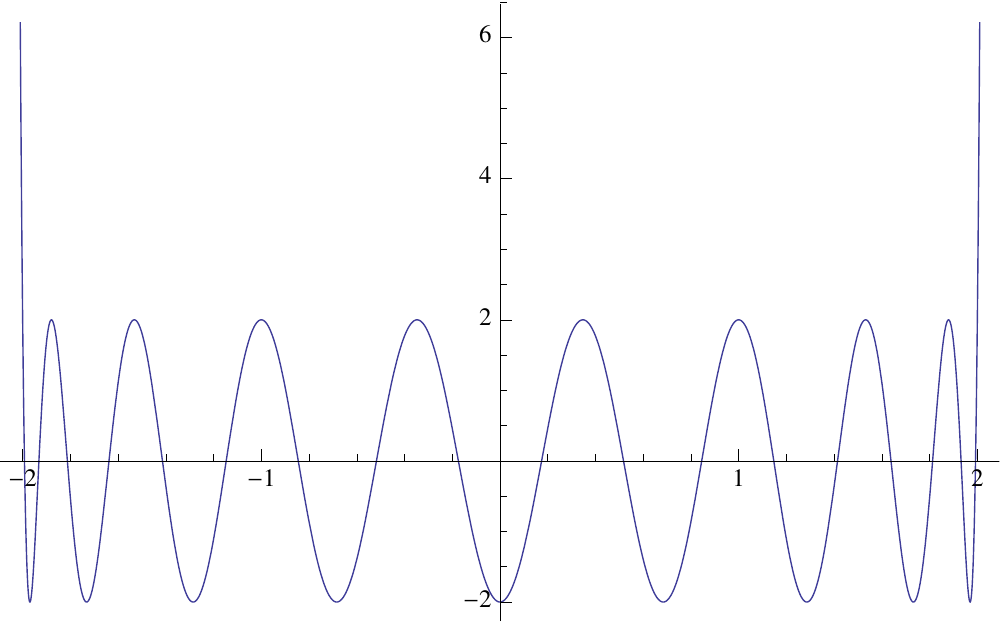}
\caption{  $P_{q=18}(E)$ }\label{fig02}
 \end{center}
\end{figure}
 
 Setting either $k_y=0$ or $\pi/(2q)$ in (\ref{zero}) one  obtains
  \be
  P_q(E) = \prod_{r= 1}^{q } \left(E -2\cos\left (\frac{2 \pi r}{q} \right)\right) +2 = \prod_{r=1}^{q}
   \left (E-2\cos\left({\displaystyle \pi\over 2 q}+{2\pi r \over q} \right)\right) =
  2  \cos \left(q \arccos\left({E\over 2}\right)\right)
  \label{P}
  \ee

 From (\ref{bibi}, \ref{zero}) the eigenenergy equation is 
  \be\label{energy} P_{q}(E)=e^{iqk_x}+2\cos(qk_y)\ee
  and, using (\ref{P}), one finally gets  
\be
 E_q(r) = 2 \cos\left[\frac{\arccos[ e^{i qk_x }/2 +\cos (q k_y)]}{q} +\frac{2\pi r}{q}\right],\quad{\rm with}\quad r = 1, 2, \ldots, q.
\label{en} \ee
or, equivalently,
\bea
 E_q(r) &=& \left( {e^{i qk_x }/2 +\cos (q k_y)+\sqrt{(e^{i qk_x }/2 +\cos (q k_y))^2-1}}\right)^{1/q}\nonumber\\&+&\left( {e^{i qk_x }/2 +\cos (q k_y)-\sqrt{(e^{i qk_x }/2 +\cos (q k_y))^2-1}}\right)^{1/q},\nonumber\\& \quad{\rm with}&\quad r = 1, 2, \ldots, q.
\label{enbis} \eea
 (\ref{en}) or (\ref{enbis})  is  the exact spectrum in the  commensurate case with arbitrary
boundary conditions. 
For  a given  such energy, which satisfyes (\ref{bibi}), iterating  (\ref{wf}) from $m=0$
\be
 \Phi_{k} = \prod_{r= 0}^{k-1 } (E -2\cos (k_y+2\pi{p\over q} r ))\Phi_{0}
  \label{wfbis}
 \ee
gives the exact eigenfunction $\Phi_{k} $ at any lattice site $k=1,2,\ldots$ in terms of a  state $\Phi_{0}$ which has to be defined by normalisation considerations.

 One has seen that  in the absence of a magnetic field $\gamma=0$, or equivalently $q=1$, the spectrum  (\ref{en}) is real when $e^{ik_x}=\pm 1$ 
\be P_{1}(E)=E=e^{ik_x}+2\cos k_y\to E_{q=1}= \pm 1 +2\cos k_y\;,\ee
i.e. (\ref{plusmoin}).
In the presence of a magnetic field $\gamma=2\pi {p\over q} $  with $q>1$, the spectrum  (\ref{en}) has  real eigenvalues  only  for $ e^{i qk_x } =\pm 1$. Setting  $e^{iqk_x}=+1 $  corresponds to periodic boundary conditions  and
$e^{iqk_x}=-1$ to anti-periodic boundary conditions on a $q-$cells lattice.

Consider for definiteness periodic boundary conditions.
In the region $| 1/2+\cos (q k_y)| < 1$
 the spectrum consists of $q$ real eigenvalues $|E_q| <2$, symmetric around  $0$ for even $q$,    and  it contains  $[q/2 +1]$ bands.  At the points where $| 1/2+\cos (q k_y)| = 1$, the spectrum is degenerate with largest eigenvalue $2$ (and by symmetry lowest eigenvalue $-2$ for even $q$).
 In the region $| 1/2+\cos (q k_y) | > 1$  there are only   $1$ or $2$  (for odd/even $q$ respectively) real eigenvalues  $|E_q| \gtrsim 2$,  and    $q-1 $ or $q-2$   complex eigenvalues   (see Fig.~3 and Fig.~4).  
\begin{figure}[tbph]
  \begin{center}
  \includegraphics[width=3.0in]{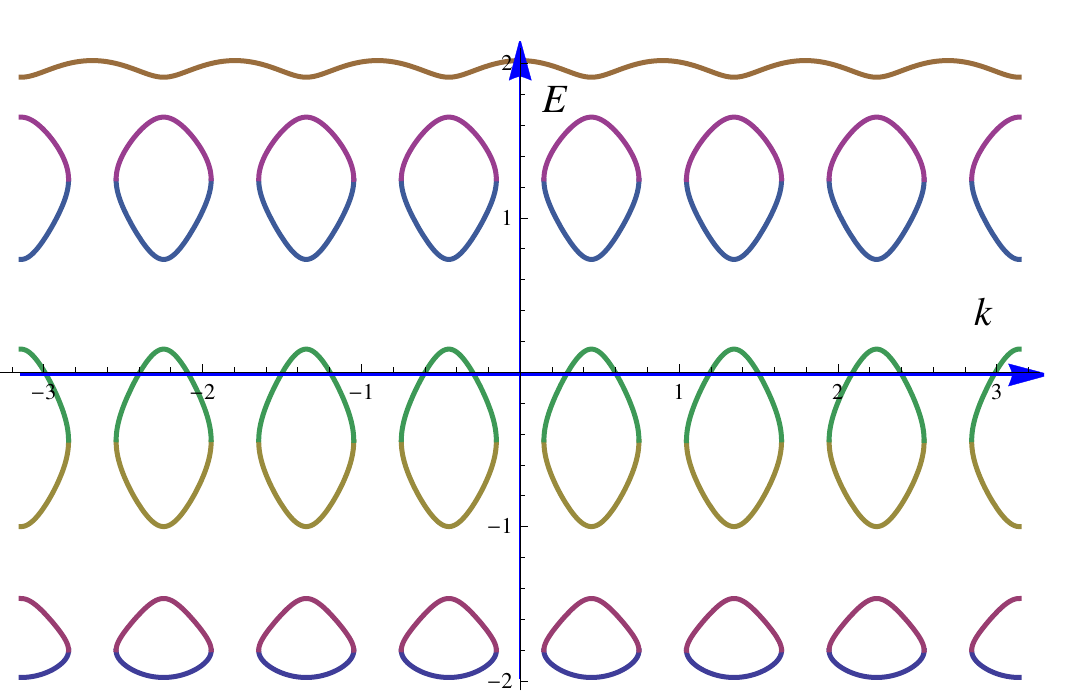}\label{fig03}
\caption{ The spectrum $E_{q=7}$ for periodic boundary conditions: on the horizontal axis $k_y\in[-\pi,\pi]$, on the vertical axis the energy.}
 \end{center}
\end{figure}
\begin{figure}[tbph]
 \begin{center}
 \includegraphics[width=3.0in]{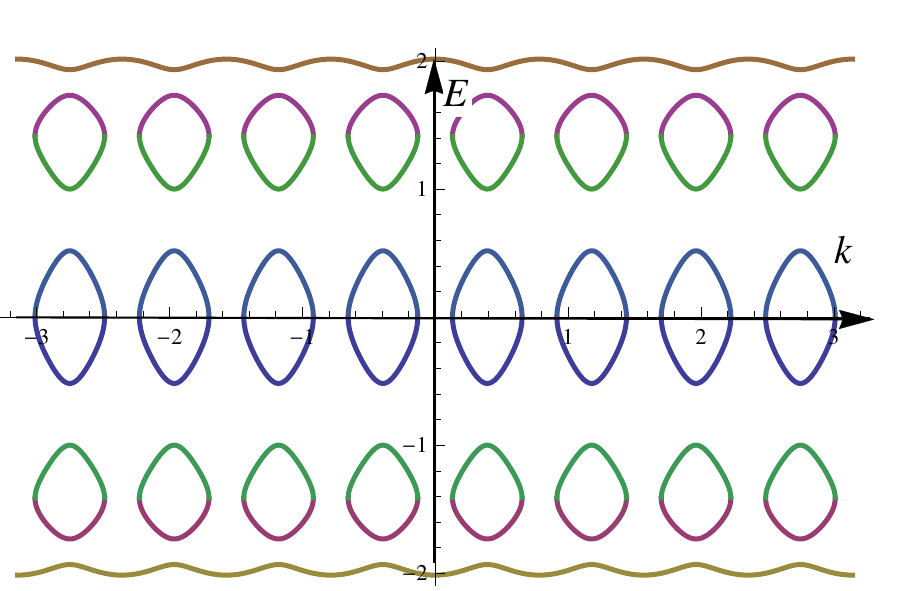}\label{fig04}
\caption{ The spectrum $E_{q=8}$ for periodic boundary conditions: on the horizontal axis $k_y\in[-\pi,\pi]$, on the vertical axis the energy.}
 \end{center}
\end{figure}
Finally, for  large $q$,  the density of  states with real energy $\rho_q (E)$  converges to the 1d tight-binding density of state
  \be\rho_q (E)   \propto  \frac{1}{\sqrt{4 - E^2}}\;.\ee
  
 The  edges\footnote{As a remark, the edges  can be easily retrieved from    the zeroes of $P_q$ in  (\ref{P}), 
\be 2\cos\left({\displaystyle \pi\over 2 q}+{2\pi r \over q} \right)\quad{\rm with}\quad r=1,2,\ldots,q\;.\ee
For example when $q=2^i$,  one gets for the zeroes
\be\bigg\{\pm\sqrt{2\pm\ldots \sqrt{2\pm\sqrt{2\pm\sqrt{2}}}}\bigg\}\label{zeroes}\ee 
where the $\dots$ means iterating $i$ times $\pm\sqrt{2}$.
The edges of the bands follow by replacing  the innermost $2$ by a $1$ and adjoining at both ends 
$\pm\sqrt{2+\ldots\sqrt{2+\sqrt{2+\sqrt{5}}}}$:
for example when  $i=5$ 
\bea&&\bigg\{\pm\sqrt{2\pm\sqrt{2\pm\sqrt{2\pm\sqrt{2\pm\sqrt{2}}}}}\bigg\}\to\nonumber\\&&
\bigg\{\pm\sqrt{2\pm\sqrt{2\pm\sqrt{2\pm\sqrt{2\pm\sqrt{1}}}}},\;\pm\sqrt{2+\sqrt{2+\sqrt{2+\sqrt{2+\sqrt{5}}}}}\bigg\}\;.\label{edges}\eea} of the  bands are  given by  the real values of the energy (\ref{en}) at  $k_y=0$ and $k_y=\pi/q$ (not ordered)
\bea &&  {\rm  for} \;q\;{\rm even:}\; \{2\cos\left({\displaystyle 2\pi\over 3 q}+{2\pi r \over q} \right)\;{\rm with}\;  r=1,2,\ldots,q,\; \pm 2 \cos\left( \frac{{\rm arccos}(3/2)}{q}\right)\} ,\nonumber\\
 &&{\rm  for} \;q\;{\rm odd:}\;\{2\cos\left({\displaystyle 2\pi\over 3 q}+{2\pi r \over q} \right) \;{\rm with}\; r=1,2,\ldots,q,\; 2 \cos\left( \frac{{\rm arccos}(3/2)}{q}\right)\}.
\eea
A plot of the edge spectrum with periodic boundary conditions for $q=1,2,\ldots,100$ and  $p\in[0,q]$ is displayed in Fig.~5.

\begin{figure}
\begin{center}
\hspace{0cm}
\vspace{0cm}
\includegraphics[scale=1]{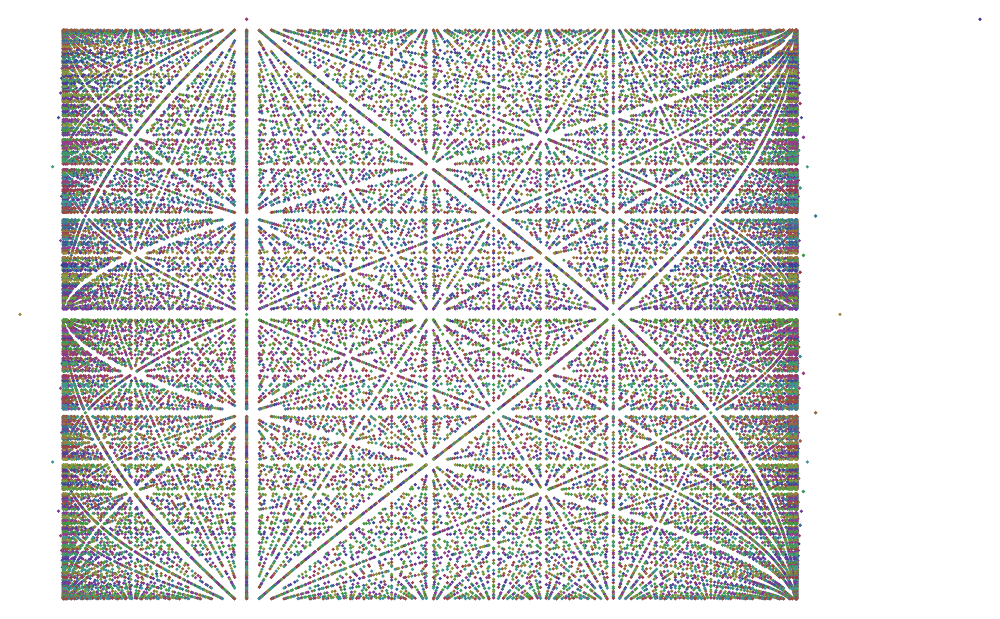}\label{figure}
\caption{ The edge spectrum for periodic boundary conditions:  $q=1,2,\ldots,55$,
 on the horizontal axis the edges and on the vertical axis $p/q\in[0,1]$.}
\end{center}
\end{figure}

The sum of the lengths of the gaps   can be easily  computed to be
$4/(1 + 2 \cos (2\pi/3q))$ for $q$  even, and   $2\cos(\pi/3q)-2/(1 + 2 \cos(\pi /3q))$ for  $q$ odd.
In the limit   $q\to\infty$, it saturates  to $4/3$ ($4/3_{+}$ when $q$ even, and    $4/3_{-}$ when $q$ odd), whereas the sum 
 of the  lengths of the bands saturates to $8/3$ ($8/3_{-}$ for $q$ even, and    $8/3_{+}$ for $q$ odd),   and the largest and  smallest eigenvalues   converge to $\pm 2$. 
A plot of the  band  spectrum as a function of $1/q$ for  $q=1,2,\ldots,100$
 is displayed  in Fig.~6. 

\begin{figure}[tbph]
\begin{center}
\includegraphics[width=4.0in]{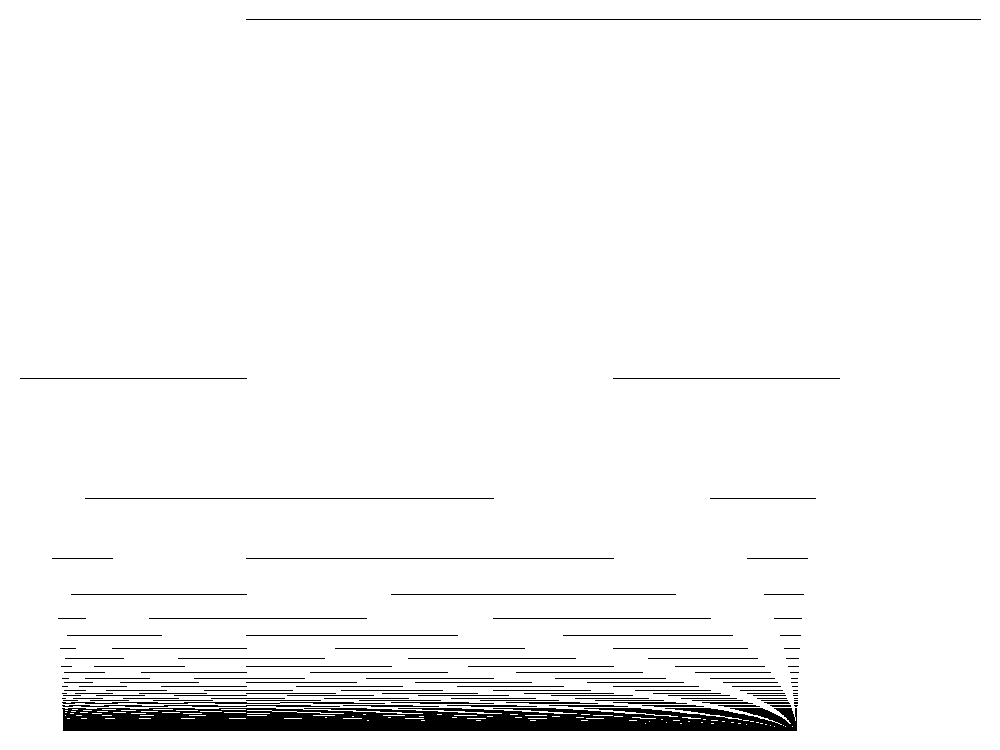}
\caption{   The band spectrum for periodic boundary conditions:  $q=1,2,\ldots,100$,  on the horizontal axis the bands and on the vertical axis $1/q$.}\label{spectr1}
 \end{center}
\end{figure}

Note that in the incommensurate case
when the flux is irrational the period has to be taken equal to the length $L$ of the lattice. For periodic boundary conditions,  one gets from (\ref{wf}) the eigenenergy equation 
\be
\Phi_L = \Phi_0   \Leftrightarrow  \prod_{r = 0}^{L-1} (E - 2 \cos ( \gamma r + k _y)) =  1
\ee
where  the sites  $m = 0$ and $m=L$  are respectively  on the left and right  sides of the lattice.
A numerical analysis shows that  for an  arbitrary wave vector $k_y$ the spectrum 
is a mixture of real and complex eigenvalues, see for example Fig.~7.
 \begin{figure}[tbph] \label{incom}
   \begin{center}
   \includegraphics[width=3.0in]{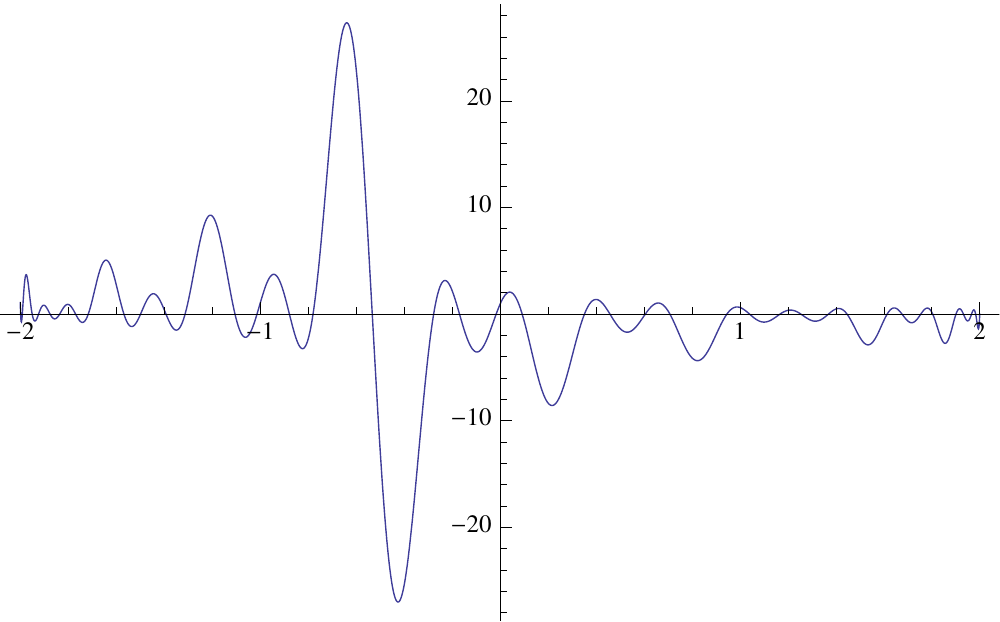}
\caption{  $\prod_{r = 0}^{L-1} (E - 2 \cos ( \gamma r + k _y))$ for $L=40$, $\gamma = 2 \pi/\sqrt{2}, k_y=0 $}
 \end{center}
\end{figure}

\section{Relation  of   $Tr  H(\gamma)^N$   to the  probability distribution generating function $Z_{N, M}({\bf q})$}

In the commensurate case $\gamma=2\pi p/q\to 2\pi /q$ the trace of the Hamiltonian (\ref{h2}) over all eigenvalues -including the complex ones- 
\be
{\rm Tr} H_{\gamma=2\pi /q}^N \equiv \int_{-\pi}^{\pi} \frac{d k_y}{2\pi} \sum_{r= 1}^{q} E_q (r)^N 
\ee
should be related to $Z_{N, M}({\bf q})$ in (\ref{solutbis}) via the 
mapping ${\bf q}=e^{i\gamma}=e^{i2\pi /q}$.  Clearly
   $q$-cells periodic boundary conditions on the quantum spectrum  should imply, for the random walks, summations over
$M$   with the same periodicity.

\subsection{Periodic boundary conditions $e^{iqk_x}=+1$}\label{per}

\noindent One can easily check that, when $q$ is even, for any even integer $N$ 
\be  {\rm Tr}H_{2\pi/  q}^{N}=\sum_{M=0}^{[N/q]}Z_{N,qM}(e^{i2\pi /q})\label{ouf} \ee
whereas for  any odd integer $N$, trivially,
\be{\rm Tr}H_{2\pi/  q}^{N}=0\ee

\noindent and when $q$ is odd
for any even integer $N$ 
\be \label{qoddneven} {\rm Tr}H_{2\pi/  q}^{N}=\sum_{M=0}^{[N/(2q)]}Z_{N,2Mq}(e^{i2\pi /q}) \ee
 and for any odd integer $N$
\be \label{qoddnodd} {\rm Tr}H_{2\pi/  q}^{N}=\sum_{M=0}^{[(N-q)/(2q)]}Z_{N,(1+2M)q}(e^{i2\pi /q}) \ee
 \subsection{Anti-periodic boundary conditions $e^{iqk_x}=-1$}\label{antiper}

   \noindent  Similarly when  $q$ is even 
   for any even integer $N$  
   \be \label{alexios} {\rm Tr}H_{2\pi/q}^{N}=\sum_{M=0}^{[N/q]}Z_{N,Mq}(e^{i2\pi /q})(-1)^M \ee
  whereas for  any odd integer $N$
   \be  {\rm Tr}H_{2\pi/  q}^{N}=0\ee
  \noindent   and when $q$ is odd
    for  any odd even integer $N$  
\be \label{qoddnevenbis} {\rm Tr}H_{2\pi/  q}^{N}=\sum_{M=0}^{[N/(2q)]}Z_{N,2Mq}(e^{i2\pi /q}) \ee
  \noindent and for  any odd integer $N$ 
\be \label{qoddnoddbis} {\rm Tr}H_{2\pi/  q}^{N}=-\sum_{M=0}^{[(N-q)/(2q)]}Z_{N,(1+2M)q}(e^{i2\pi /q}) \ee
 
\noindent Note that when $\gamma=0$, i.e. $q=1$,  (\ref{qoddneven}, \ref{qoddnodd}, \ref{qoddnevenbis}, \ref{qoddnoddbis}) correctly reproduce (\ref{even}, \ref{odd}).
  
 \subsection{General boundary conditions  $e^{iqk_x}\ne \pm 1$}\label{k_x}
\noindent  So far, with periodic or antiperiodic boundary conditions,  one has satisfied that the  spectrum be real when $\gamma=0$ and that, eventhough   part of the spectrum is complex when $\gamma\ne 0$, the traces in Sections \ref{per} and \ref{antiper} be as well  real.  
 One can drop the reality requirement and go a step further  by considering general boundary conditions,  $e^{iqk_x}\ne \pm 1$.
 
 \noindent  In the case  $q$ even, from (\ref{ouf}, \ref{alexios})  one infers  
  when $N$ is even
 \be  \label{1}{\rm Tr}H_{2\pi/  q}^{N}=\sum_{M=0}^{[N/q]}Z_{N,Mq}(e^{i2\pi /q})e^{iMqk_x} \ee
  and  when $N$ is odd
 \be   \label{2}{\rm Tr}H_{2\pi/  q}^{N}=0\ee
 \noindent In the  case $q$ odd, from (\ref{qoddneven}, \ref{qoddnodd}, \ref{qoddnevenbis}, \ref{qoddnoddbis}) , one arrives 
  when $N$ is even at
\be \label{3} {\rm Tr}H_{2\pi/  q}^{N}=\sum_{M=0}^{[N/(2q)]}Z_{N,2Mq}(e^{i2\pi /q})e^{i2Mqk_x } \ee
\noindent and when $N$  is odd at
\be \label{4} {\rm Tr}H_{2\pi/  q}^{N}=\sum_{M=0}^{[(N-q)/(2q)]}Z_{N,(1+2M)q}(e^{i2\pi /q})e^{i(1+2M)qk_x} \ee
Note  that the RHS of (\ref{1}, \ref{2}, \ref{3}, \ref{4}) are unchanged if the argument $e^{i2\pi /q}$ in $Z_{N,M}$ is replaced by $e^{i2\pi p/q}$, as it should.  Note also that these  equations narrow down to the same and unique equation (\ref{1}) provided that one decides that each time a $Z_{N,M}$ appearing in the sum has entries  with different parities, it should be considered as zero: $N$ and $M$ have indeed to have the same parity for the classical random walk to actually exist.

\section{Discussion and Conclusion}

Motivated by exact results for the probability distribution of the algebraic area of biased random walks, we introduced a  quantum mechanical  2d  lattice  model  of charged particles  with an anisotropic    hopping coupled to a perpendicular uniform magnetic field.  An exact  solution in the case of a commensurate magnetic flux per unit cell was found with explicit expressions for the  eigenvalues  and eigenfunctions  spectrum.
The Hamiltonian  is non hermitian due to the anisotropic hopping -the absence of right $\to$ left
  hopping: it  describes a quantum Hall effect setting of  enforced left $\to$ right motions for  charged particles
  in a perpendicular magnetic field.  As a result,  
  the Hamiltonian has non real eigenvalues which imply a damping of the corresponding states. 
  
  We 
  explicitly mapped the     biased classical random system  on  the non hermitian  Hofstadter-like quantum model.   When the magnetic flux per unit cell is rational, periodicity  on the lattice allows to relate the biased length $N$  random walks algebraic area probability generating functions  to the traces of the $N^{\rm th}$ power of  the non hermitian Hofstadter-like quantum Hamiltonian. The identities~(\ref{1}, \ref{2}, \ref{3},  \ref{4})  are the main results obtained so far\footnote{We stress that these identities have been checked numerically  for various $N$ up to $N=20$ and $q$ up to $q=10$ but we did not  derive them analytically.}: they do encapsulate the mapping between both models.  Conversely, it would be interesting to see  at a possible geometric interpretation  of  the RHS of  (\ref{1}, \ref{2}, \ref{3}, \ref{4})  for the biased random  walks themselves.

  It is interesting to note that one can recover the hermiticity  of the Hamiltonian 
   by considering  the system with a uniform  constant current $J$, so that 
   the Hamiltonian becomes $ H - \lambda J$, where $\lambda $ is a Lagrange   multiplier,
   $ J =i \sum (\psi^+_{n+ 1} \psi_n -   \psi^+_{n} \psi_{n+1})$ and $H =  \sum \psi^+_{n+ 1} \psi_n + \ldots$.
   Then, for the  appropriate choice    $\lambda = -i/2$, the resulting Hamiltonian becomes hermitian.

{\bf Acknowledgements:}  S.O. would like to thank Stefan Mashkevich for collaboration in the early stages of this work (section \ref{just})  and Alexios Polychronakos for suggesting to look at  general boundary conditions (section \ref{k_x}).  He would also like to thank Alain Comtet and Christophe Texier for interesting discussions.  The help of Etienne Werly has also been precious  for  some technical details and  the plots of the spectra.
\end{document}